\documentclass[useAMS]{mn2e} 
\usepackage{epsfig,lscape} 
\usepackage{graphicx}

\DeclareGraphicsExtensions{.eps,.eps.gz,.epsi}

\newcommand\ao{Appl. Opt.} 
\newcommand\procspie{Proc. SPIE}%
\newcommand\aj{AJ}%
\newcommand\apj{ApJ}%
\newcommand\apjl{ApJL}%
\newcommand\aap{A\&A}%
\newcommand\pasp{PASP}%
\newcommand\mnras{MNRAS}%
\newcommand\araa{ARA\&A}%
\newcommand\apjs{ApJS}%
\newcommand\pasj{PASJ}%

\newcommand{\teff}{\mbox{$T_{\rm eff}$}} 
\newcommand{\logg}{{\rm{log}~$g$}}
\newcommand{\feh}{{\rm [Fe/H]}} 
\newcommand{\ebv}{$E(B-V)$}
\newcommand{\rab}{$R(a-b)$} 
\newcommand{\rv}{$R(V)$}
\newcommand{\ra}{$R(a)$}

\begin{document}

\title[Empirical extinction coefficients from the far-UV to the mid-IR]{Empirical extinction
coefficients for the GALEX, SDSS, 2MASS and WISE passbands}

\author[Yuan, Liu \& Xiang]{H. B. Yuan$^{1}$\thanks{LAMOST
Fellow}\thanks{E-mail: yuanhb4861@pku.edu.cn}, X. W. Liu$^{1,2}$ \& M. S. Xiang$^{2}$\\ \\ 
$1$ Kavli Institute for Astronomy and Astrophysics, Peking University, Beijing 100871, P. R. China \\ 
$2$ Department of Astronomy, Peking University, Beijing 100871, P. R. China}

\date{Received:}

\pagerange{\pageref{firstpage}--\pageref{lastpage}} \pubyear{2013}

\maketitle

\label{firstpage}

\begin{abstract}{
Using the "standard pair" technique of paring stars of almost nil and high extinction
but otherwise of almost identical stellar parameters from the Sloan Digital Sky
Survey (SDSS), and combing the SDSS, Galaxy Evolution Explore (GALEX), Two Micro All Sky Survey (2MASS) 
and Wide-field Infrared Survey Explorer (WISE) photometry ranging from the far ultraviolet (UV) to
the mid-infrared (mid-IR), we have measured dust reddening in the $FUV-NUV, NUV-u, u-g,
g-r, r-i, i-z, z-J, J-H, H-Ks, Ks-W1$ and $W1-W2$ colors for thousands of
Galactic stars.  The measurements, together with the \ebv~values given by Schlegel et al. (1998), 
allow us to derive the observed, model-free reddening coefficients for those colors.
The results are compared with previous measurements and the predictions of a variety of 
Galactic reddening laws. We find that 1) The dust reddening map of Schlegel et
al. (1998) over-estimates \ebv~by about 14 per cent, consistent with the recent work of
Schlafly et al. (2010) and Schlafly \& Finkbeiner (2011); 2) After
accounted for the differences in reddening normalization, the newly deduced reddening coefficients for colors $FUV-NUV,
NUV-u, u-g, g-r, r-i, i-z, z-J, J-H$ and $H-Ks$ differ by respectively
$-$1640\%, 15.5\%, 12.6\%, $-$0.8\%, 3.4\%, $-$0.7\%, 3.5\%, 2.5\% and 1.4\% from 
the predictions of Fitzpatrick reddening law (Fitzpatrick 1999) 
for an assumed total-to-selective extinction ratio \rv~=~3.1,
and by respectively $-$1730\%, 13.0\%, 8.1\%, 10.0\%, 8.0\%, $-$13.5\%,
$-$1.7\%, $-$6.7\% and $-$17.1\% from the predictions 
of CCM reddening law (Cardelli et al. 1989); and 3)  All the new reddening
coefficients, except those for $NUV-u$ and $u-g$,  prefer the \rv~=~3.1 Fitzpatrick
reddening law rather than the \rv~=~3.1 CCM and O'Donnell (O'Donnell 1994) reddening laws. Using the $Ks$-band extinction
coefficient predicted by the \rv~=~3.1 Fitzpatrick law and the
observed reddening coefficients, we have deduced new extinction
coefficients for the $FUV, NUV, u, g, r, i, z, J, H, W1$ and $W2$ passbands.
We recommend that the new reddening and extinction coefficients should be used in
the future and an update of the Fitzpatrick reddening law in the UV is probably necessary.  
We stress however that the $FUV$- and $NUV$-band coefficients should be used 
with caution given their relatively large measurement uncertainties.
Finally, potential applications of the "standard pair" 
technique  with the LAMOST Galactic surveys are discussed.} 
\end{abstract}


\section{Introduction} Dust grains produce extinction and reddening of stellar light from
the ultraviolet (UV) to the infrared (IR) (Draine 2003).
Accurate determination of reddening to a star is vital for reliable derivation of 
its basic stellar parameters, such as distance, effective temperature and intrinsic
spectral energy distribution (SED).

Using the {\it IRAS} and DIRBE data, Schlegel et al. (1998, hereafter SFD) has
generated a whole sky 2D dust-reddening map of \ebv~based on the dust thermal emission. 
The map has been widely used to correct for extinction and reddening of
extra galactic as well as Galactic sources.
However, the SFD dust reddening map delivers the total amount of
reddening along a sightline at a spatial resolution of about 6 arcmin, 
thus it may have over-estimated the real values for Galactic sources.  
The extinction of a given passband
$a$ and the color excess (reddening) of a given color $a-b$ are usually estimated by
$A(a)$\,=\,\ra\,$\times$\,\ebv~and $E(a-b)$\,=\,\rab\,$\times$\,\ebv,
respectively.  Here \ra, defined as extinction in $a$ band relative to \ebv, and
$R(a-b)=R(a)-R(b)$, defined as reddening in $a-b$ color relative to \ebv, are the $a$ band
extinction and $a-b$ color reddening coefficients, respectively.  Values of
\ra~ and \rab~ are usually computed from an extinction curve.  
Cardelli et al. (1989) obtained an \rv-dependent Galactic extinction law over the wavelength range 0.125$\mu$m $\le$ $\lambda \le$ 3.5$\mu$m.
O'Donnell (1994) derived a slightly different \rv-dependent extinction law in the optical and near-UV 
(0.303$\mu$m $\le \lambda \le$ 0.909$\mu$m). In 1999, Fitzpatrick presented a new  average Galactic 
extinction curve from the IR through the UV (0.1$\mu$m $\le \lambda \le$ 3.5$\mu$m).
The \rv~=~3.1 extinction law of O'Donnell (1994) and that of Cardelli et al. (1989) 
for the wavelength range outside the O'Donnell law were used by SFD to 
derive the extinction coefficients for a large number of passbands.  Various
studies have however shown that SFD may have over-estimated extinction in both the high- and
low-extinction regimes (Arce \& Goodman 1999; Chen et al. 1999;
Dobashi et al. 2005; Yasuda et al. 2007; Rowles \& Froebrich 2009).

The Sloan Digital Sky Survey (SDSS; York et al. 2000) has provided uniform and
contiguous imaging data of about one-third of the sky in the $u,g,r,i$ and $z$ bands
(DR8; Aihara et al. 2011).  The "blue tip" of the stellar locus provides a
viable color standard to measure and study reddening.  Using the "blue tip"
method, Schlafly et al. (2010, hereafter S10) find that Fitzpatrick reddening law
(Fitzpatrick 1999) of \rv~=~3.1 is preferred to those of O'Donnell (1994) and Cardelli et
al. (1989) adopted by SFD.  They also find that the SFD 
map traces the dust well, but it overestimates \ebv~by 14 per cent and reddening in
$u-g, g-r, r-i$ and $i-z$ colors by up to 40 per cent. 

The SDSS DR8 also delivers low-resolution spectra for
over 500,000 stars, along with stellar parameters deduced from the Sloan
Extension for Galactic Understanding and Exploration (SEGUE; Yanny et al. 2009)
Stellar Parameter Pipeline (SSPP, Lee et al. 2008a,b; Allende Prieto et al.
2008; Lee et al. 2011; Smolinski et al. 2011).  A fraction of the spectroscopic
targets are towards high extinction sightlines.  Using intrinsic colors 
from the MARCS grid of model atmospheres (Gustafsson et
al. 2008), Schlafly \& Finkbeiner (2011, hereafter SF11) estimates values of reddening 
for a sample of reddened stars based on the SDSS spectra, yielding results consistent with S10.

Given the large number of stars with SDSS spectra and precise stellar parameters, 
for any target star that has been significantly reddened by interstellar dust
grains, one can always find another essentially unreddened star or group of stars of
almost identical stellar parameters (effective temperature, surface gravity and metallicity) 
that can serve as the control sample that provides the intrinsic colors for the target star.
Reddening of the target in a given color can then easily be calculated by 
comparing its observed color and intrinsic value, with the latter derived from its control stars.  
In a previous paper, we have used the  "standard pair" technique (Stecher, 1965; Massa, Savage \& Fitzpatrick, 1983)
 to detect and measure diffuse interstellar bands (DIBs) towards sightlines
of a large sample of reddened stars (Yuan \& Liu 2012). In the current work,
the same technique is used to measure and study dust reddening and extinction from the
UV to the mid-IR by combining data from the 
SDSS and other photometric surveys, including the Galaxy Evolution Explorer (GALEX;
Martin et al. 2005) in the UV, the Two Micro All Sky Survey (2MASS; Skrutskie et al. 2006) in the near-IR
and the Wide-field Infrared Survey Explorer (WISE; Wright et al. 2010) in the mid-IR.
Model-free reddening and extinction coefficients are obtained
and used to test the calibration of SFD and various extinction laws.

The paper is organized as follows.  In Section 2, we introduce the data sets
and method used to measure the reddening in different colors.  The reddening
coefficients for different colors and extinction coefficients for different
passbands are presented in Section 3 and compared with previous studies 
and the predictions of a variety of Galactic reddening laws. 
The results are discussed in Section 4, along with the potential applications 
of the "standard pair" technique in several on-going and forth-coming 
large scale spectroscopic surveys. A brief summary then follows in Section 5.

\section{Data and Method} 

Extinction curves are usually measured using the "standard pair" technique (Stecher, 1965; Massa, Savage \& Fitzpatrick, 1983) by
comparing the photometric and/or spectrophotometric measurements of two stars of the same spectral type,
one has negligible foreground dust while the other heavily reddened.
Comparison of the SEDs of the two stars, together with the assumption that the dust
extinction goes to zero at very long wavelengths, allows one to determine the
extinction $A_{\lambda} = 2.5 \times log(F_{\lambda0}/F_{\lambda})$ as a
function of wavelength $\lambda$, where $F_{\lambda}$ is the observed flux and
$F_{\lambda0}$ is the flux in the absence of extinction. The method has
been used to measure extinction curves for many sightlines, in many cases over
a wavelength range extending from the vacuum UV to the near-IR.
We use a similar "standard pair" method to measure reddening and then derive reddening
coefficient of a given color.  This method requires a target sample of highly reddened
stars and a control sample of unreddened or of extremely low reddening stars of matching spectral types.
Control stars of matching spectral type are used to estimate the intrinsic colors of a target star.
In order to maximize the numbers of target and control stars for different
passbands, different selection criteria have been used.  This Section describes selections
of target and control stars for measuring reddening coefficients of the SDSS,
GALEX, 2MASS and WISE passbands.

\begin{figure*} 
\centering 
\includegraphics[width=180mm]{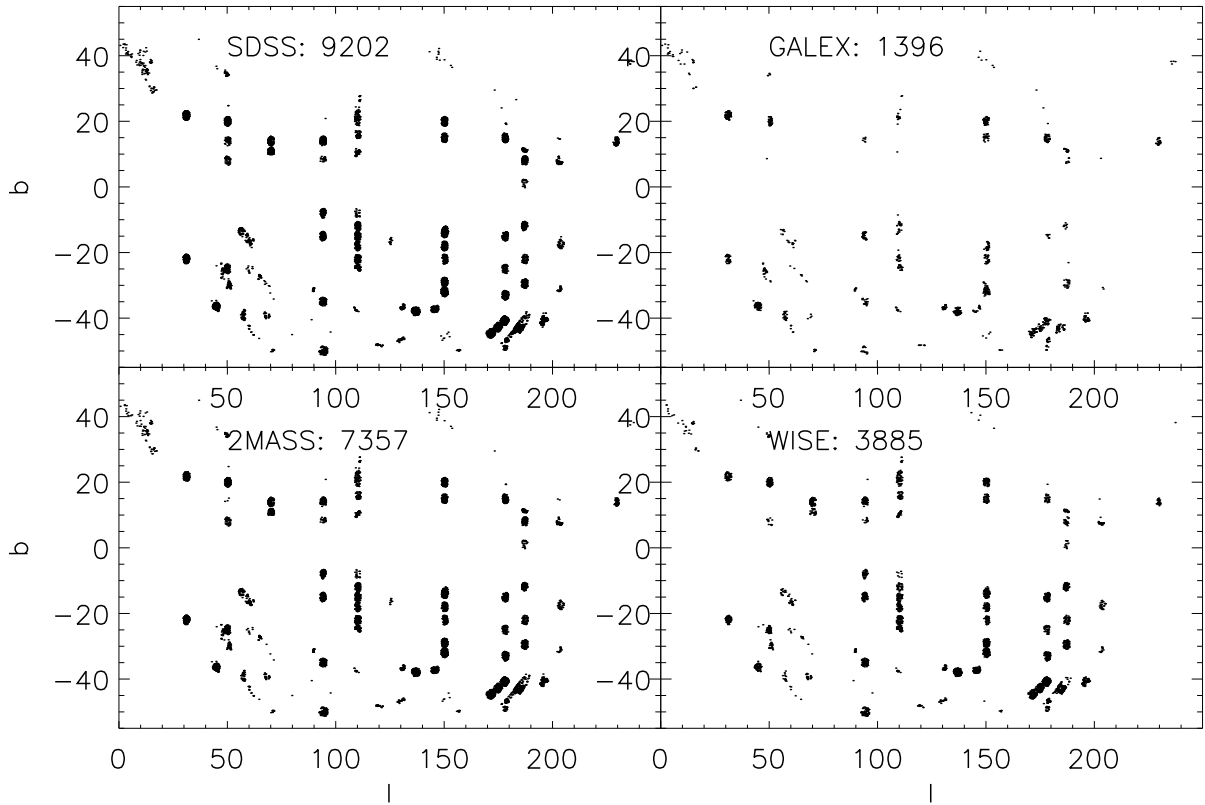}
\caption{Spatial distributions of target samples for measuring reddening
coefficients of the SDSS (top left), GALEX (top right), 2MASS (bottom left) and
WISE (bottom right) passbands. The total numbers of stars of the individual samples are also marked.} \label{}
\end{figure*}

\begin{figure*} 
\centering 
\includegraphics[width=180mm]{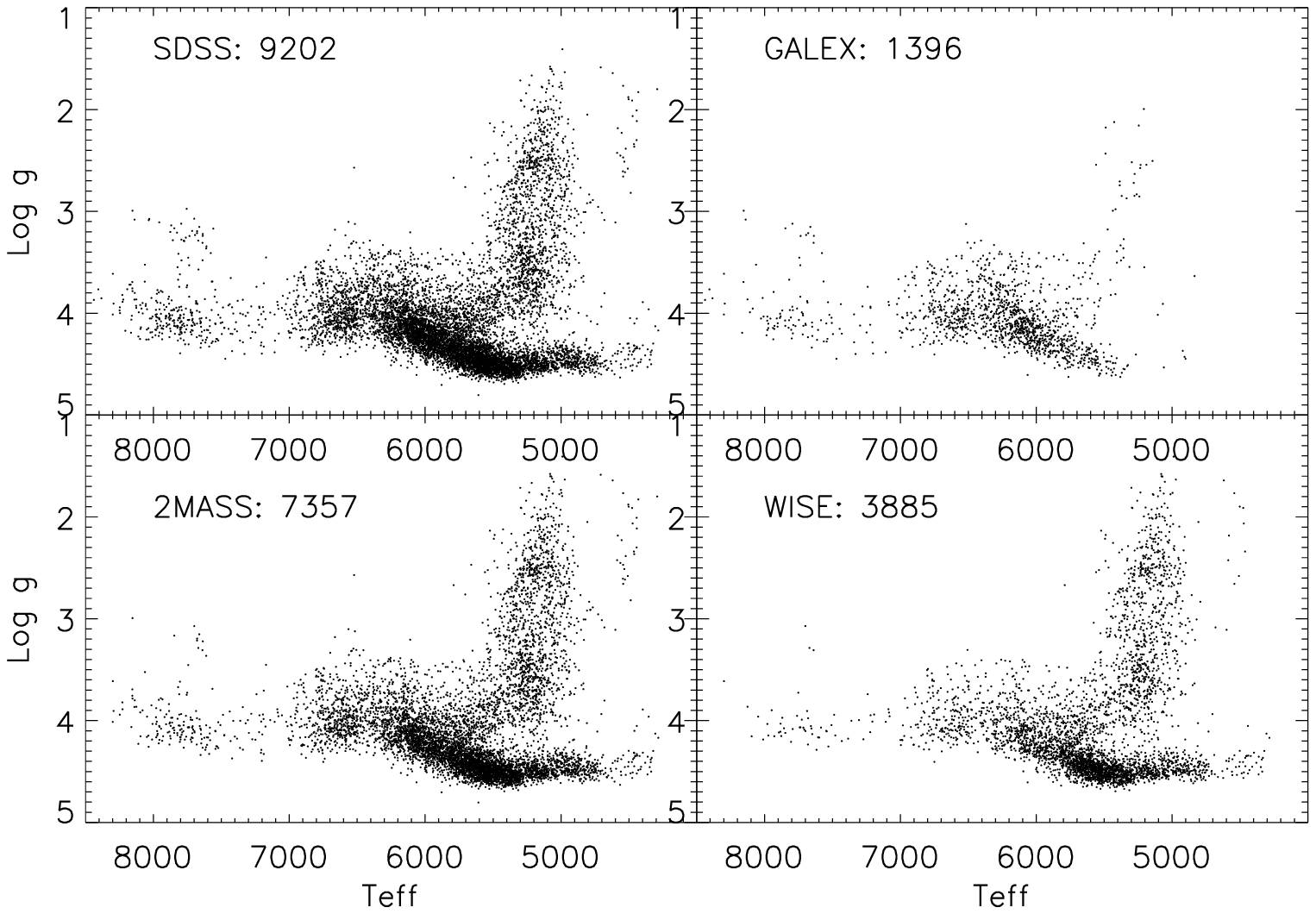}
\caption{$\teff$ versus \logg~distributions of target samples for measuring reddening
coefficients of the SDSS (top left), GALEX (top right), 2MASS (bottom left) and
WISE (bottom right) passbands. The total numbers of stars of the individual samples are also marked.} \label{}
\end{figure*}

\subsection{Data for the SDSS passbands} 
Both the target and control
samples are selected from the SDSS Data Release 7 (DR7; Abazajian et al. 2009).
SDSS DR7 provides accurate photometry of about 100 million stars in $u,g,r,i$
and $z$ bands and spectra of more than 300,000 stars.  We select target stars
as those with a line-of-sight extinction \ebv~$>$ 0.1, with a spectrum of
signal-to-noise ratio S/N $>$ 40 and having basic stellar parameters, effective temperature \teff, surface gravity
\logg~and metallicity [Fe/H], well
determined with corresponding errors smaller than 100 K, 0.2 dex and 0.1 dex, respectively.
Values of \ebv~used here are from SFD.
Values of  \teff, \logg~and [Fe/H] are from the SSPP.
Note that SSPP estimates \teff, \logg~and [Fe/H] of each star using a variety 
of methods.  Some of the methods are photometry-based and thus have some sensitivity to
reddening corrections, while the others, such as the NGS1, ki13, ANNRR and ANNSR
methods, do not rely on photometry.  To avoid uncertainties introduced by reddening
corrections, we use the mean values of \teff, \logg~and [Fe/H] derived from the
NGS1, ki13, ANNRR and ANNSR methods only for the purpose of the current work. 
However, the errors estimated by SSPP are still used.
Typical uncertainties given by SSPP for \teff, \logg~and [Fe/H] are 180 K, 0.25 dex and 0.23
dex, respectively, dominated by the systematic errors (Schlesinger et al. 2010;
Smolinski et al. 2011).  In this work, we are mainly interested in the
relative ranking of stars in the \teff, \logg~and \feh~ parameter spaces, thus 
systematic uncertainties are not important.  In total, 9,202 target stars
are selected. Their spatial and $\teff$ versus \logg~distributions are shown in the top left panels of
Fig.\,1 and Fig.\,2, respectively. It is seen that the target stars are mainly composed of 
FGK dwarfs, with a small fraction of them being A dwarfs and KM giants. 
These targets are from different SDSS spectroscopic plates, so they are spatially clustered by plate.

For the control sample, we select stars with a line-of-sight extinction \ebv~$<$
0.03 and a spectrum of S/N $>$ 20.  The err cuts on \teff, \logg~and
\feh~are the same as target stars.  In total, 50,053 stars are selected.

\subsection{Data for the GALEX passbands} 
GALEX (Martin et al. 2005) is a space telescope providing imaging in the far-UV
(1344--1786{\AA}, centered at 1528{\AA}) and near-UV (1771--2831{\AA}, centered
at 2271{\AA}) bands, with a 6--8 arcsec angular resolution (80 per cent encircled energy)
and 1 arcsec astrometry.  GALEX is carrying out a number of surveys
of different sizes of sky coverage and detection depths.  The catalogs of unique UV sources
from two GALEX's surveys: AIS (the All-Sky Imaging Survey of depths of AB magnitudes of 19.9 and 20.8 in
far- and near-UV bands, respectively) and MIS (the Medium-depth Imaging Survey of depths of 22.6 and 22.7, respectively) from
GALEX fifth data release have been matched to the SDSS DR7 catalogs by Bianchi et
al. (2011), using a match radius of 3.0 arcsec.

The target and control samples for the GALEX passbands are selected from the corresponding
samples for the SDSS passbands by requiring that the sources are well detected in the two
GALEX bands with photometric errors [error(FUV) and error(NUV)] smaller than 0.2.  
Given the relatively low angular resolution of 
GALEX, a GALEX source may have multiple SDSS matches.
Sources with multiple matches within a search radius of 4.2 arcsec are also
removed. In total, 1,396 targets and 16,405 control stars are collected.
The spatial and $\teff$ versus \logg~distributions of the targets are shown in the top right panels of Fig.\,1 
and Fig.\,2, respectively. The target stars are also spatially clustered, and most of them are FG dwarfs.

\subsection{Data for the 2MASS passbands} 
2MASS (Skrutskie et al. 2006) has made uniformly-calibrated imaging observations of the whole
sky in the $J$ (1.24 $\mu$m), $H$ (1.66 $\mu$m) and $Ks$ (2.16 $\mu$m)
near-IR bands.  The 2MASS Point Source Catalog contains positions and
photometry for 470,992,970 objects, and is complete down to $J=15.8$, $H=15.1$
and $Ks=14.3$ mag.

The target and control samples for the 2MASS passbands are also selected from
the corresponding samples for the SDSS passbands by requiring that the sources are well detected and
having photometric errors smaller than 0.1 mag in all the three 2MASS bands.
In total, 7,357 targets and 34,548 control stars are collected.  
The spatial and $\teff$ versus \logg~distributions of the targets, which are very similar to those of the targets for
the SDSS passbands, are plotted in the  bottom left panels of Fig.\,1
and Fig.\,2, respectively.

\subsection{Data for the WISE passbands} 
The satellite WISE (Wright et al. 2010) has imaged the whole sky in four bands at 3.4, 4.6,
12, and 22 $\mu$m (named $W1$, $W2$, $W3$ and $W4$, respectively)  with a corresponding angular resolution of 6.1, 6.4, 6.5
and 12.0 arcsec.  The WISE Source Catalog
contains positions and photometry for over 563 million point-like and resolved
objects.  The positions are calibrated against the 2MASS, achieving an accuracy
of  $\sim$200 mas on each axis with respect to the 2MASS reference frame for
sources of S/N's better than 40.  Photometry is performed using techniques of point source
profile-fitting and multi-aperture photometry, achieving 5$\sigma$ photometric
sensitivities of 0.068, 0.098, 0.86 and 5.4 mJy (equivalent to 16.6, 15.6, 11.3 and 8.0 Vega
mag) at 3.4, 4.6, 12 and 22 $\mu$m, respectively, in unconfused regions in the ecliptic plane.
Given the low sensitivities of the $W3$ and $W4$ bands and the poor angular resolution of the $W4$ band, 
the extinction coefficients for the two bands are not measured in this work. We focus only
on the $W1$ and $W2$ bands.

The target and control samples for the WISE $W1$ and $W2$ passbands are selected
from the corresponding samples for the 2MASS passbands, requiring that the sources are well detected in 
the $W1$ and $W2$ bands and have photometric errors smaller than 0.05 mag in both
bands. The samples for the 2MASS passbands are matched to the WISE
sources with a match radius of 3.0 arcsec.  Considering the low angular
resolutions of WISE, WISE sources with multiple matches within a search radius
of 10.0 arcsec are excluded.  Sources that are affected by known artifacts or
are likely variables are also removed. In total, 3,885 targets and 10,842 control
stars are collected. 
The spatial and $\teff$ versus \logg~distributions of the targets are plotted in the  bottom right panels of Fig.\,1
and Fig.\,2, respectively. Most of the targets are GK dwarfs and giants.

\subsection{Method} The "standard pair" technique used in this work is
very similar to the template subtraction method used by Yuan \& Liu (2012) to
detect and measure DIBs in the SDSS spectra.  For each star in
the target sample, its control stars are selected from the control sample as
those having values of \teff, \logg~and \feh~that differ from those of
the target by smaller than 50 K, 0.25 dex and 0.1 dex, respectively. The
reddening of the target in a given color is measured as the difference between
the observed and intrinsic colors. The latter is derived assuming that the
intrinsic colors of the target and its control stars vary linearly with \teff,
\logg~and \feh. The assumption is valid considering the small ranges of values of
\teff, \logg~and \feh~being considered. The control stars are dereddened using an initial
set of reddening coefficients and \ebv~values from SFD.  Then a new set of reddening
coefficients is derived by comparing the estimates of reddening relative to \ebv~
for the target sample.  Iterations are carried out till the derived set of 
reddening coefficients is consistent with the one used for dereddening 
the control sample.
  
After obtaining reddening coefficients for colors of two adjacent bands, the
extinction coefficients for all bands can be computed by assuming an extinction coefficient value for a
given passband. 

\section{Results}
 
\begin{table*} \begin{minipage}[]{160mm} \centering 
\caption{\rab~for different colors.} 
\label{} 
\begin{tabular}{|l|l|l|l|l|l|l|l|l|l|} \hline Color
&This work$^{a}$ &This work$^{b}$   &Fitzpatrick$^{c, d}$ &CCM$^{d, e}$ & O'Donnell$^{d, f}$  & SF11$^{g}$   & S10$^{h}$ &  SFD$^{i}$          \\\hline 
$FUV-NUV$ &$-$2.35$\pm$0.58 &$-$2.69$\pm$0.50 & 0.164  & 0.154  &             & &  &                                 \\ \hline 
$NUV-u$ &2.85$\pm$0.070&2.71$\pm$0.22   & 2.406         & 2.460  &             & &                & \\ \hline 
$u-g$ &1.08$\pm$0.010   &1.04$\pm$0.018  & 0.945        &0.984  &0.950 & 0.94$\pm$0.02      & 1.01$\pm$0.10 &  1.362                \\ \hline 
$g-r$ &0.99$\pm$0.015   &0.99$\pm$0.015  & 0.999       &0.901  &0.936             & 0.98$\pm$0.02     & 1.01$\pm$0.08 &  1.042                  \\ \hline 
$r-i$ &0.60$\pm$0.010  & 0.60$\pm$0.011 & 0.582       &0.557   &0.514             & 0.55$\pm$0.01      &0.57$\pm$0.05 &  0.665                  \\ \hline 
$i-z$ &0.43$\pm$0.004   &0.43$\pm$0.005  & 0.426     &0.496    &0.513             & 0.44$\pm$0.01       &0.45$\pm$0.05  &  0.607                \\ \hline
$z-J$ &0.56$\pm$0.011   &0.56$\pm$0.013  & 0.544     &0.554    &             & & &                                       \\ \hline 
$J-H$ &0.26$\pm$0.005 &0.26$\pm$0.011  & 0.254       &0.279     &             & &             & \\ \hline 
$H-Ks$   & 0.16$\pm$0.006  & 0.16$\pm$0.005 & 0.158   &0.188     & & &             &                                         \\ \hline 
$Ks-W1$  & 0.12$\pm$0.008  & 0.12$\pm$0.010 & 0.120    &0.149     &             & & &                                          \\ \hline 
$W1-W2$  & 0.026$\pm$0.004 & 0.036$\pm$0.007 & 0.063  &0.075     &             & &             & \\ \hline 
\end{tabular} 
\begin{description} 
\item[$^a$]  \rab~derived by fitting $E(a-b)$ versus $E(g-r)$ diagram, assuming $R(g-r)$~=~0.99.  
\item[$^b$]  \rab~derived by fitting $E(a-b)$ against \ebv~from SFD.  
\item[$^c$]  Predictions by an \rv~=~3.1 Fitzpatrick extinction law at \ebv~=~0.4 for a 7,000 K source
spectrum, assuming that SFD over-estimates \ebv~by 14 per cent. 
\item[$^d$] The filter curves for GALEX, SDSS, 2MASS and WISE passbands are from 
Morrissey et al. (2004), the SDSS DR7 website, Cohen et al. (2003) and Wright et al. (2010), respectively.
\item[$^e$]  Predictions by an \rv~=~3.1 CCM extinction law at \ebv~=~0.4 for
a 7,000 K source spectrum, assuming that SFD over-estimates \ebv~by 14 per cent.  
\item[$^f$]  Predictions by an \rv~=~3.1 O'Donnell extinction law at \ebv~=~0.4 for
a 7,000 K source spectrum, assuming that SFD over-estimates \ebv~by 14 per cent.  
\item[$^g$]  \rab~derived from fitting SFD by SF11 using stars with spectra without
applying zero-point offsets.  
\item[$^h$]  \rab~derived from fitting SFD by S10 with the "Blue tip" method.  
\item[$^i$]  Original SFD prescription.
\end{description} \end{minipage} \end{table*}

\subsection{$R(u-g)$, $R(g-r)$, $R(r-i)$ and $R(i-z)$} 

We first determine the reddening coefficients for the $u-g$, $g-r$, $r-i$ and $i-z$ colors for the sample of the SDSS
passbands.  The left panels in Fig.\,3 show reddening (i.e. color excess, defined as the difference of the 
observed and intrinsic colors) $E(u-g)$, $E(r-i)$ and $E(i-z)$  
versus $E(g-r)$. All those derived are independent on possible uncertainties in the SFD map.  The right panels in
Fig.\,3 show the same set of reddening plus $E(g-r)$ plotted against \ebv~given by
SFD. The black pluses denote results from the individual target stars, 
whereas the big red pluses represent median values of individual data points 
grouped into 8 bins in the X-axis, with a bin size of 0.1 magnitude.
A 3-$\sigma$ clipping has been applied in calculating the medians.
The red lines represent linear regressions passing through the origin of the median values, with each point carrying 
equal weight.  
As already mentioned, the SFD map delivers the total amount of extinction integrated along a given line-of-sight to infinite, 
has a limited spatial resolution about 6 arcmin and fails at low Galactic latitudes ($|b| \le 5\degr$), thus the \ebv values 
from SFD map may not represent the true values of the targets. However, as to be discussed in Section\,4,
these effects are not important for the targets in this work.

The bottom right panel of Fig.\,3 yields an $R(g-r)$ value of 0.99$\pm$0.015, 
which is consistent with those obtained by S10 and SF11, confirming their earlier findings 
that \ebv~values from SFD are over-estimated by about 14 per cent. 
The number 14 per cent is consistent with the fact that an \rv~=~3.1 Fitzpatrick law predicts $R(g-r)$~=~1.139.
Adopting the relation $E(g-r)$~=~0.99 $\times$ \ebv, values of
$R(u-g)$, $R(r-i)$ and $R(i-z)$ can then be derived from the data plotted in the left panels of
Fig.\,3. The results are found to agree well with those derived from the data plotted in the right panels.  
The values of $R(u-g)$, $R(g-r)$, $R(r-i)$ and $R(i-z)$ deduced in this work are
listed in the 2nd and 3rd columns of Table\,1. The 4th, 5th and 6th columns give predictions of 
the \rv~=~3.1 Fitzpatrick, CCM and O'Donnell extinction laws, respectively, 
assuming that SFD over-predicts \ebv~by 14 per cent.
For comparison, relations predicted by the Fitzpatrick, CCM and O'Donnell laws are
over-plotted in Fig.\,3 in purple, blue and cyan, respectively.
The predicted values are calculated by convolving a synthetic stellar spectral model
from Castelli and Kurucz (2004) of \teff~=~7,000 K, \logg~=~4.5 and \feh~=~$−1$ without and with dust extinction of \ebv~=~0.4.
The SDSS filter curves are from the SDSS DR7 website.
Values obtained by SF11, S10 and SFD are also
listed in Table\,1.

Note that the extinction and reddening coefficients predicted by the extinction 
laws do have some dependence on the source spectrum.  
For example, for a temperature range from 5,000 -- 7,000\,K that covers most targets in this work,
the variations of $R(a)$ predicted by the \rv~=~3.1 Fitzpatrick law 
(except $R(FUV)$, $R(NUV)$, $R(g)$ and $R(r)$) in this work caused by such dependence are found to be well below 0.01. 
The predicted $R(FUV)$, $R(NUV)$, $R(g)$ and $R(r)$ increase respectively by $-0.95$, 0.76, 0.069 and 0.016  when
the source temperature increases from 5,000 -- 7,000\,K.
Therefore,the dependence of \ra~ and consequently \rab~on source spectrum can be safely ignored in most cases.
Similarly, the extinction and reddening coefficients also depend on the amount of extinction.  
The predicted $R(FUV)$, $R(NUV)$, $R(u)$, $R(g)$ and $R(r)$ decrease respectively by 0.02, 0.45, 0.015, 0.047 and 0.016  
when \ebv~increases from 0.2 to 1.0.
However, for other bands considered in this work, the
corresponding variations for \ebv~between 0.2 and 1.0 are well below 0.01. 
Thus again it is reasonable to assume constant
extinction and reddening coefficients for the purpose of this work.

As in the case of $R(g-r)$, values of $R(r-i)$ and $R(i-z)$ yielded by our data 
are also consistent with those of S10 and SF11. A comparison of values of $R(g-r)$, $R(r-i)$ and
$R(i-z)$ of the current work with those predicted by different extinction laws
prefers the \rv~=~3.1 Fitzpatrick reddening law to those of CCM and O'Donnell,
again consistent with the findings of S10 and S11.  However, $R(u-g)$ is worst predicted.
The value of $R(u-g)$ in the 2nd column of Table\,1 derived in this work, is 
respectively 7 and 15 per cent higher than those from S10 and S11, and is respectively
14, 10 and 14 per cent higher than those predicted by the
\rv~=~3.1 Fitzpatrick, CCM and O'Donnell laws.  Based on a color-color fit, SF11
obtains a relation $E(u-g)$ = 1.01 $\times$ $E(g-r)$.  Checking the data plotted in the
left panel of Fig.\,7 of SF11, we find that their above relation under-predicts
$E(u-g)$/$E(g-r)$ by about 10 per cent at high extinctions.  The same
problem also exists in the left panel of Fig.\,16 of S10.  The top panels of
Fig.\,3 also show a similar trend that $R(u-g)$ tends to be
larger at higher extinctions.  Thus all the data, those of this work and of SF11 and S10 are consistent
and the differences in the derived
reddening coefficients, such as $R(u-g)$ and $R(i-z)$, are probably largely caused by
the different fitting procedures.

\begin{figure*} \centering \includegraphics[width=180mm]{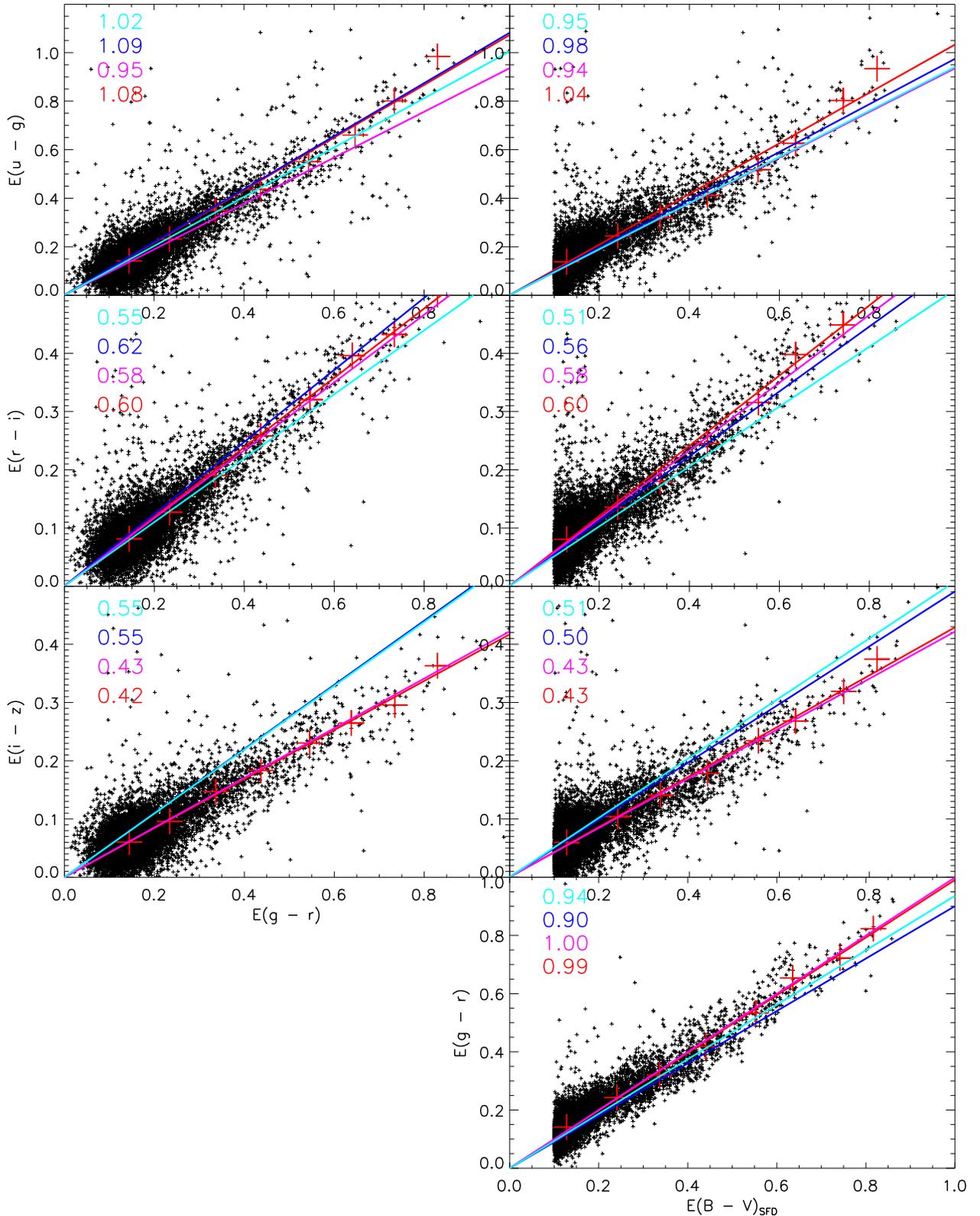}
\caption{Reddening coefficients of the $u-g$, $g-r$, $r-i$ and $i-z$ colors
deduced using the samples for the SDSS passbands.  
\emph{Left:} Reddening of $u-g$, $r-i$ and $i-z$ colors versus that of $g-r$.  
\emph{Right:} Reddening of
$u-g$, $g-r$, $r-i$ and $i-z$ colors versus \ebv~from SFD.  Black pluses denote 
data deduced for individual stars. Large, red pluses represent 
median values by binning the data points into eight groups with a bin size of 0.1 in the X-axis. 
The red lines are linear regressions passing through the origin to the red pluses,
with each plus carrying equal weight.  For comparison, relations predicted
by the Fitzpatrick, CCM and O'Donnell laws of \rv~=~3.1 are over-plotted in purple,
blue and cyan, respectively. 
} \label{} \end{figure*}

\subsection{$R(NUV-u)$ and $R(FUV-NUV)$} 
Fig.\,4 shows reddening of $NUV-u$
and $FUV-NUV$ colors versus that of $g-r$ and \ebv~of the target sample for the GALEX
passbands.  The symbols and lines are similar to those in Fig.\,3. Due to the
large (even systematic) photometric uncertainties of GALEX data, the small size of the target sample,
the relatively strong sensitivity on the temperature of the targets and possibly 
stellar chromospheric activities of solar-type stars, values of
$R(NUV-u)$ and $R(FUV-NUV)$ deduced suffer large uncertainties, especially for $R(FUV-NUV)$.
However, the values derived from the data plotted in the left and right panels of Fig.\,4 
are still consistent within the error bars, as listed in Table\,1.
Relations predicted by the \rv~=~3.1  Fitzpatrick and CCM laws are also listed.
The GALEX filter curves are from Morrissey et al. (2004).
Both Fitzpatrick and CCM laws seem to have under-predicted $R(NUV-u)$ by about 12 per cent and 
over-predicted $R(FUV-NUV)$ dramatically.

\begin{figure*} \centering \includegraphics[width=180mm]{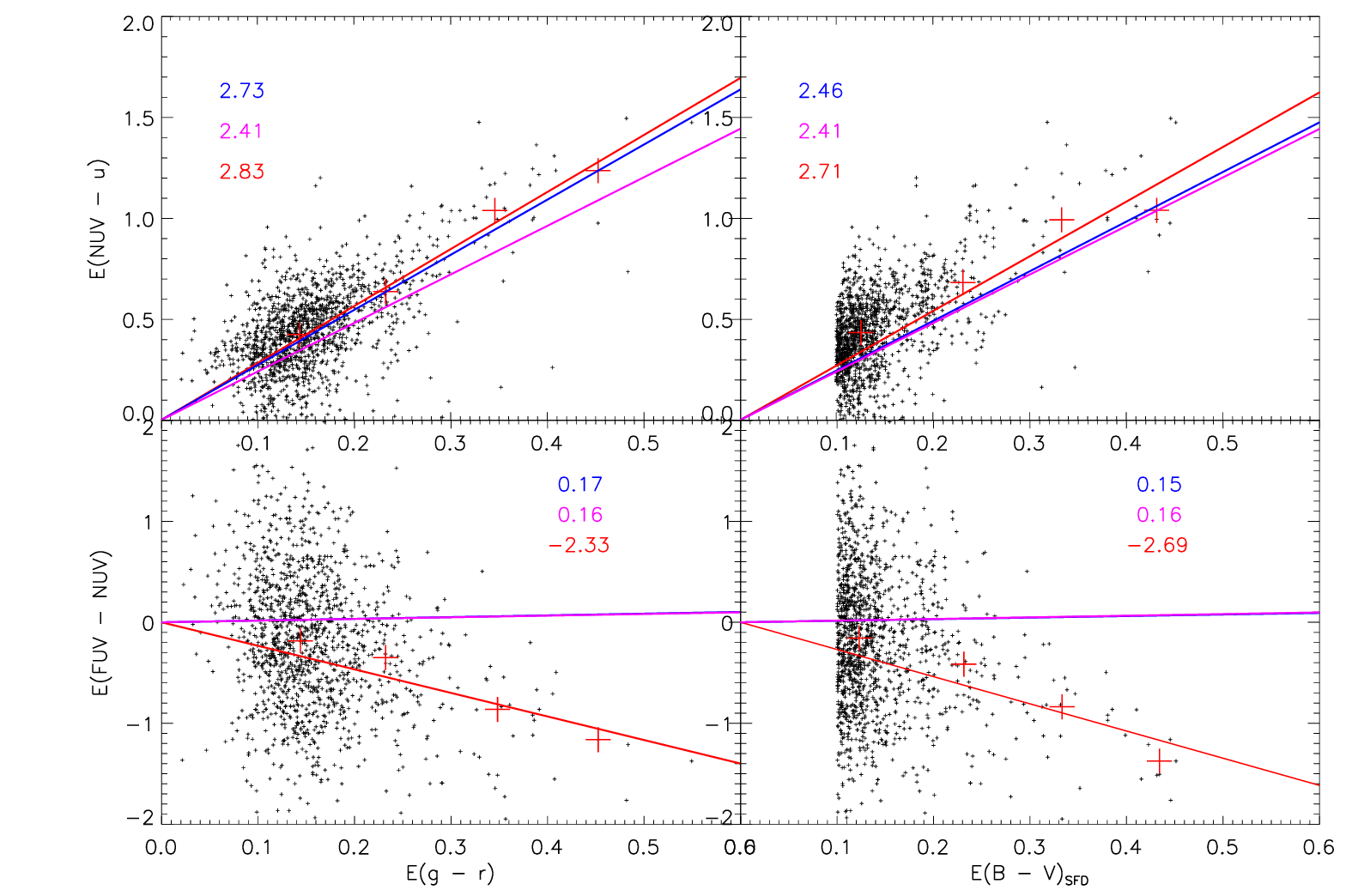}
\caption{ Reddening coefficients of the $NUV-u$ and $FUV-NUV$ colors deduced
using the samples for the GALEX passbands.  \emph{Left:} Reddening of $NUV-u$
and $FUV-NUV$ colors versus that of $g-r$.  \emph{Right:} Reddening of $NUV-u$
and $FUV-NUV$ colors versus \ebv~ from SFD.  The symbols and lines are similar to
those in Fig.\,3.  }

\label{} \end{figure*}

\subsection{$R(z-J)$, $R(J-H)$ and $R(H-Ks)$} 
Fig.\,5 shows reddening of
$z-J$, $J-H$ and $H-Ks$ colors versus that of $g-r$ and \ebv~of the target sample for
the 2MASS passbands.  The symbols and lines are similar to those in Fig.\,3.
Again due to the relatively large photometric uncertainties of 2MASS for faint sources, the scatters in Fig.\,5
are larger than those in Fig.\,3.  Values of $R(z-J)$, $R(J-H)$ and
$R(H-Ks)$ deduced are compared to the predictions of the \rv~=~3.1  Fitzpatrick
and CCM laws in Table\,1. The 2MASS filter curves are from Cohen et al. (2003).

Values of $R(z-J)$, $R(J-H)$ and $R(H-Ks)$ obtained from data plotted in the left panels of
Fig.\,5, consistent with those deduced from data plotted in the right panels, differs by  3.5, 2.5
and 1.4 per cent respectively from those predicted by the \rv~=~3.1 Fitzpatrick law,
and by $-$1.7, $-$6.7 and $-$17.1 per cent respectively from those  by the
\rv~=~3.1 CCM law.  Our measured values of $R(z-J)$,
$R(J-H)$ and $R(H-Ks)$ prefer the \rv~=~3.1 Fitzpatrick law to that of
CCM.

\begin{figure*} \centering \includegraphics[width=180mm]{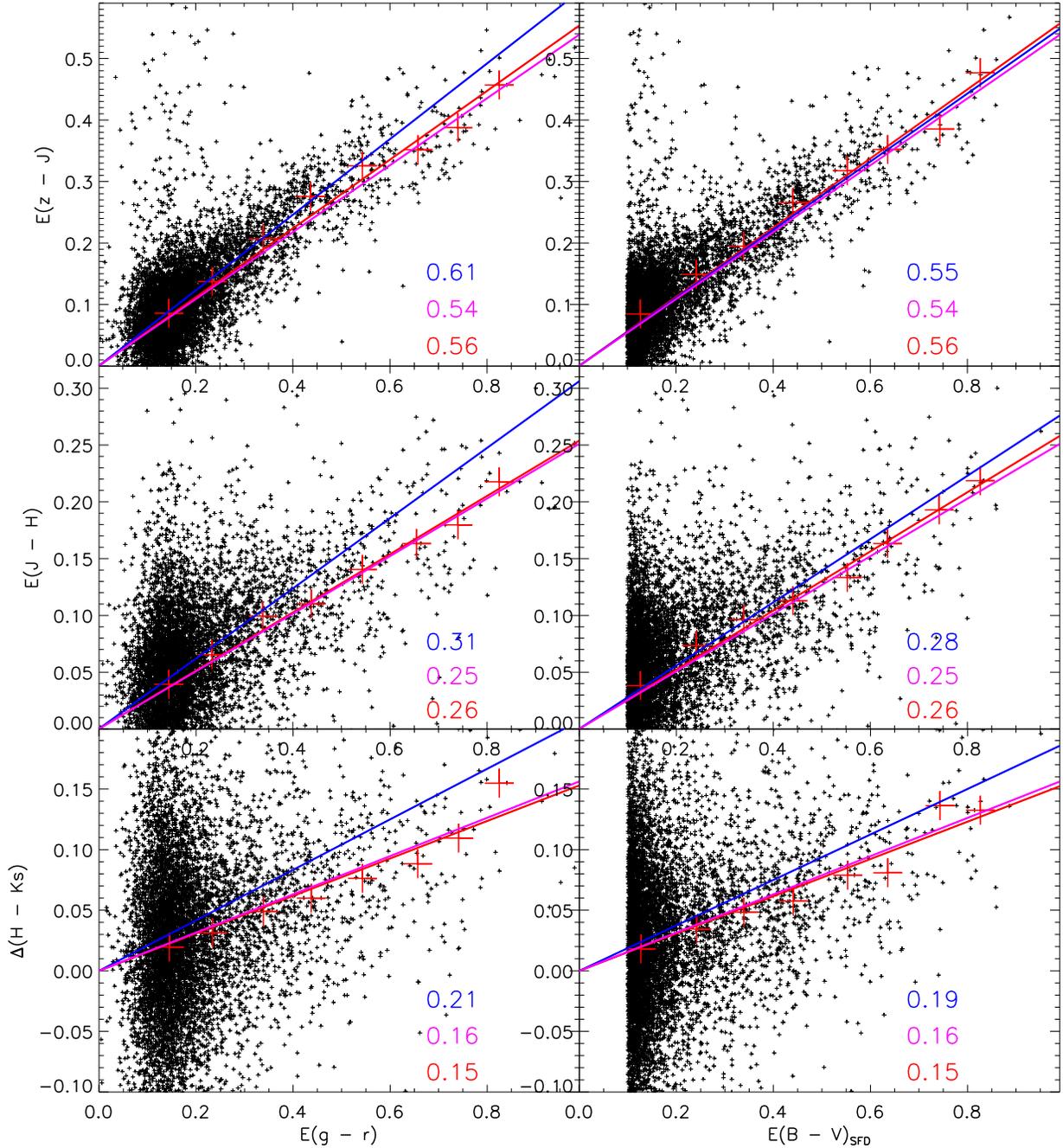}
\caption{ Reddening coefficients of the $z-J$, $J-H$ and $H-Ks$ colors deduced
using the samples for the 2MASS passbands.  \emph{Left:} Reddening of $z-J$,
$J-H$ and $H-Ks$ colors versus that of $g-r$.  \emph{Right:} Reddening of $z-J$,
$J-H$ and $H-Ks$ colors versus \ebv~ from SFD.  The symbols and lines are similar to
those in Fig.\,3.  } \label{} 
\end{figure*}

\subsection{$R(Ks-W1)$ and $R(W1-W2)$} 
Fig.\,6 shows reddening of $Ks-W1$ and
$W1-W2$ colors versus that of $g-r$ and \ebv~ of the target sample for the WISE
passbands.  The symbols and lines are similar to those in Fig.\,3.  We obtained
$R(Ks-W1)$ = 0.12$\pm$0.008 and $R(W1-W2)$ = 0.026$\pm$0.004 from data plotted in the left
panels and $R(Ks-W1)$ = 0.12$\pm$0.010 and $R(W1-W2)$ = 0.036$\pm$0.007 from those in
the right panels, respectively.  The numbers are listed in Table\,1, along with
predictions of the Fitzpatrick and CCM reddening laws.  
The WISE filter curves are from Wright et al. (2010).
Note that the Fitzpatrick law is only valid from 0.1 to 3.5 microns,
and the CCM law is only valid from 0.125 to 3.5 microns.
Beyond 3.5 micron, the extinction laws are extrapolated to calculate 
the predictions for the WISE passbands, which may cause uncorrect results. 
Again, the results prefer the \rv~=~3.1 Fitzpatrick law to that of CCM.  
Unsurprisingly, both the Fitzpatrick and CCM laws over-predict $R(W1-W2)$ significantly.
However, the observed small values of $R(W1-W2)$ are consistent with previous studies 
of mid-IR extinction laws (e.g., Gao, Jiang \& Li 2009).

\begin{figure*} \centering
\includegraphics[width=180mm]{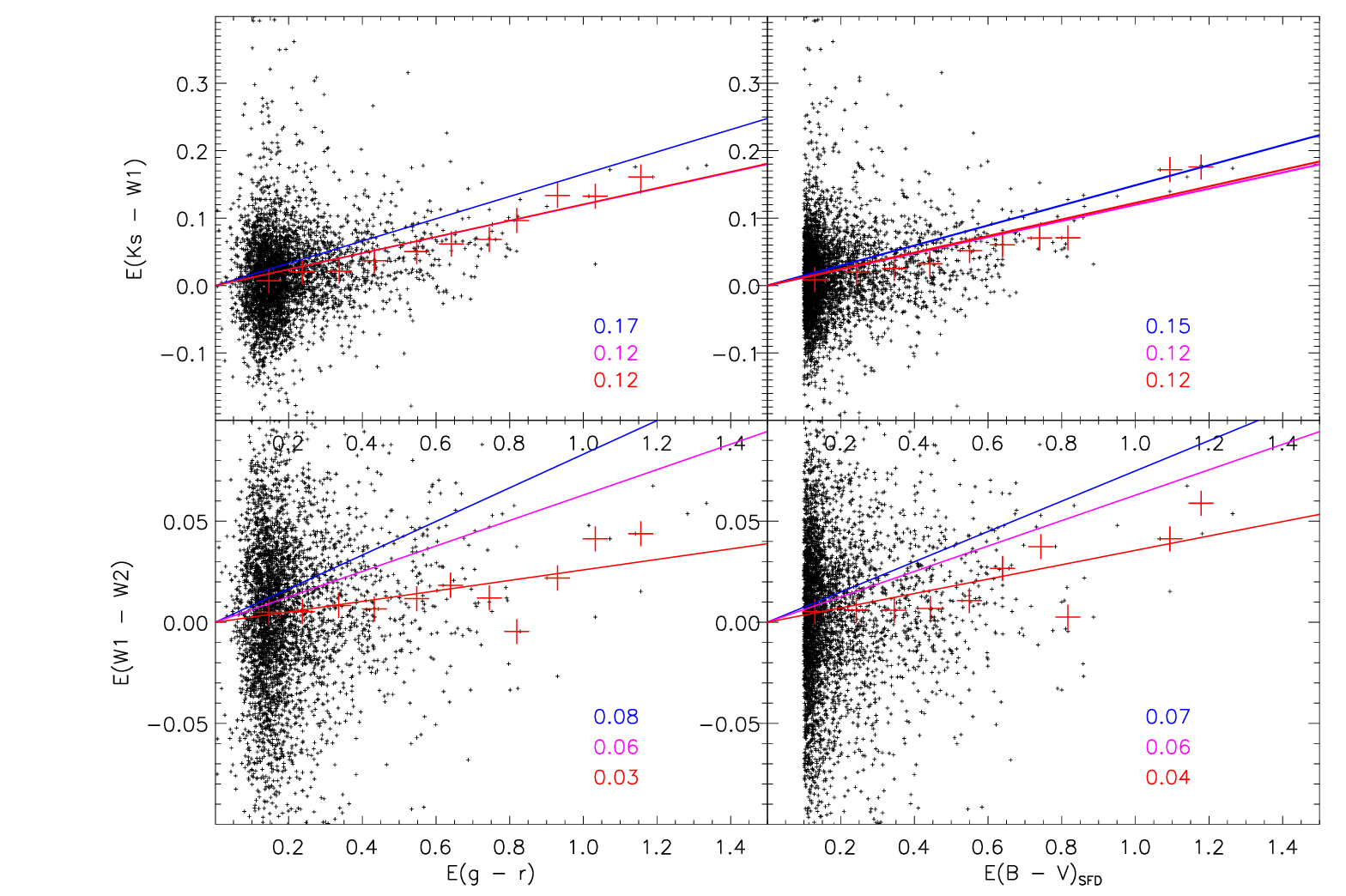} \caption{ Reddening
coefficients of the $Ks-W1$ and $W1-W2$ colors deduced using the samples for the
WISE passbands.  \emph{Left:} Reddening of $Ks-W1$ and $W1-W2$ colors versus that
of $g-r$.  \emph{Right:} Reddening of $Ks-W1$ and $W1-W2$ colors versus \ebv~ from
SFD.  The symbols and lines are similar to those in Fig.\,3.  } \label{}
\end{figure*}

\subsection {Extinction coefficients for the GALEX, SDSS,  2MASS and WISE passbands}
To derive the extinction coefficients for the GALEX, SDSS,  2MASS and WISE
passbands from the reddening coefficients for colors of adjacent bands presented above,
\ra~value for a certain band is needed. Ideally, if one of the band studied has long enough wavelength, say in the 
far-IR, then one can always assume that $R(a)$ for that band is zero.
This is however not the case for the current study.
As such, we have adopted $R(Ks)$ = 0.306, predicted by the \rv~=~3.1 Fitzpatrick law assuming that 
SFD over-predicts \ebv~by 14 per cent, as the reference point.
Here the \rv~=~3.1 Fitzpatrick law is used because it is favored by the observations in this work.
The resultant extinction coefficients for the GALEX, SDSS,
2MASS and WISE passbands are given in Table\,2.  Also listed in Table\,2 include
extinction coefficients predicted by the \rv~=~3.1 Fitzpatrick, CCM and O'Donnell
laws assuming that SFD over-predicts \ebv~by 14 per cent and those adopted in the literature for comparison.

Table\,2 shows that the two sets of \ra~values based on the reddening coefficients deduced from the data plotted 
in the left and right panels of Figs.\,3--6 agree with each other within 2.5 per cent, except
$R(FUV)$ and $R(W2)$, for which the results differ by about 10 per cent.
Table\,2 also shows the derived \ra~values overall agree better with the 
\rv~=~3.1 Fitzpatrick law than the \rv~=~3.1 CCM and O'Donnell laws.
Both the \rv~=~3.1 Fitzpatrick and CCM laws over-predict $R(FUV)$
by 40 -- 50 per cent and under-predict $R(W2)$ by 20 -- 30 per cent.  There
are also large differences between the \ra~values deduced in this work and those of
SFD, Seibert et al. (2005) and Majewski et al. (2003).

\begin{table*} \begin{minipage}[]{140mm} \centering \caption{\ra~for the GALEX, SDSS,  2MASS and WISE passbands.} \label{}
\begin{tabular}{|l|l|l|l|l|l|l|l|l|l|} \hline 
Band  &This work$^{a}$  &This work$^{b}$ & Fitzpatrick$^{c,d}$ & CCM$^{d,e}$   & O'Donnell$^{d,f}$  & SFD$^{g}$  & Seibert$^{h}$ & Majewski$^{i}$ \\ \hline 
$FUV$ &4.89$\pm$0.60     &4.37$\pm$0.54      &  6.783     &6.892   & &            &8.29           &  \\   \hline   
$NUV$ &7.24$\pm$0.08     &7.06$\pm$0.22      &  6.620     &6.738    &              &            &8.18           & \\ \hline 
$u$   &4.39$\pm$0.04     &4.35$\pm$0.04      &  4.214     &4.278    & 4.259 & 5.155        &               &  \\ \hline 
$g$   &3.30$\pm$0.03    &3.31$\pm$0.03      &  3.269     &3.294    & 3.309           & 3.793        &               &  \\ \hline 
$r$   &2.31$\pm$0.03     &2.32$\pm$0.03      & 2.270      &2.393    & 2.373           & 2.751 &               & \\ \hline 
$i$ &1.71  $\pm$0.02     &1.72$\pm$0.02      &  1.689     &1.836    & 1.859           & 2.086        & &\\ \hline 
$z$   &1.29$\pm$40.02    &1.28$\pm$0.02      &  1.261     &1.340    & 1.346          & 1.479 &    &\\ \hline 
$J$   &0.72$\pm$0.01    &0.72$\pm$0.01      &  0.717     &0.786    &  & &    &0.82\\ \hline 
$H$   &0.46$\pm$0.01     &0.46$\pm$0.01     &  0.464     &0.508    &             &             &   &0.53\\ \hline 
$Ks$  &0.306             &0.306             &0.306       &0.320    &             &             &   &0.34\\ \hline 
$W1$  &0.18 $\pm$0.01    &0.19$\pm$0.01      &  0.186     &0.171    &  &            &   &    \\ \hline 
$W2$  &0.16 $\pm$0.01   &0.15$\pm$0.01      &  0.123      &0.096    &              &            & &    \\ \hline 
\end{tabular} \begin{description} \item[$^a$]  Calculated using
$R(Ks)=0.306$ and reddening coefficients from the 2nd column of Table\,1.
\item[$^b$]  Calculated using $R(Ks)=0.306$ and reddening coefficients from
the 3rd column of Table\,1.  
\item[$^c$]  Predictions of an \rv~=~3.1
Fitzpatrick extinction law at \ebv~=~0.4 for a 7,000 K source spectrum,
assuming that SFD over-predicts the true values of \ebv~by 14 per cent.
\item[$^d$] The filter curves for GALEX, SDSS, 2MASS and WISE passbands are from 
Morrissey et al. (2004), the SDSS DR7 website, Cohen et al. (2003) and Wright et al. (2010), respectively.
\item[$^e$]  Predictions of an \rv~=~3.1 CCM extinction law at \ebv~=
0.4 for a 7,000 K source spectrum, assuming that SFD over-predicts the true values of \ebv~by 14 per cent. 
\item[$^f$]  Predictions of an \rv~=~3.1 O'Donnell extinction law at \ebv~=
0.4 for a 7,000 K source spectrum, assuming that SFD over-predicts the true values of \ebv~by 14 per cent. 
\item[$^g$]  Original SFD prescription.
\item[$^h$]  From Seibert et al.  (2005).  \item[$^i$]  From Majewski et al.
(2003).  \end{description} \end{minipage} \end{table*}

\section{Discussion} 
\begin{figure}
\includegraphics[width=90mm]{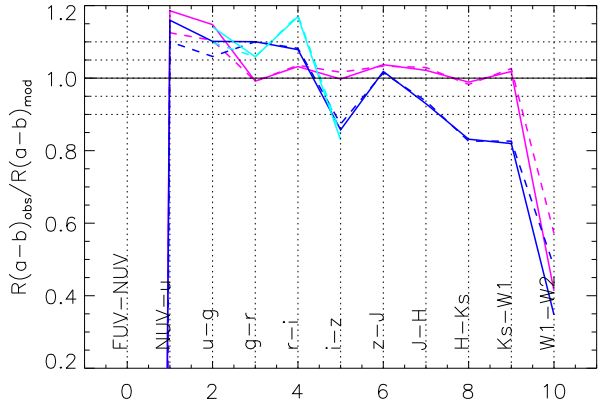}
\caption{Comparison between the measured and predicted reddening coefficients using data from Table\,1.
The purple solid and dashed lines represent the ratios of the measured reddening coefficients in the 2nd and 3rd columns to 
those predicted by the \rv~=~3.1 Fitzpatrick law in the 4th column, respectively.
The blue solid and dashed lines represent the ratios of the measured reddening coefficients in the 2nd and 3rd columns to 
those predicted by the \rv~=~3.1 CCM law in the 5th column, respectively.
The cyan solid and dashed lines represent the ratios of the measured reddening coefficients in the 2nd and 3rd columns to 
those predicted by the \rv~=~3.1 O'Donnell law in the 6th column, respectively.
}
\label{}
\end{figure}
\begin{figure}
\includegraphics[width=90mm]{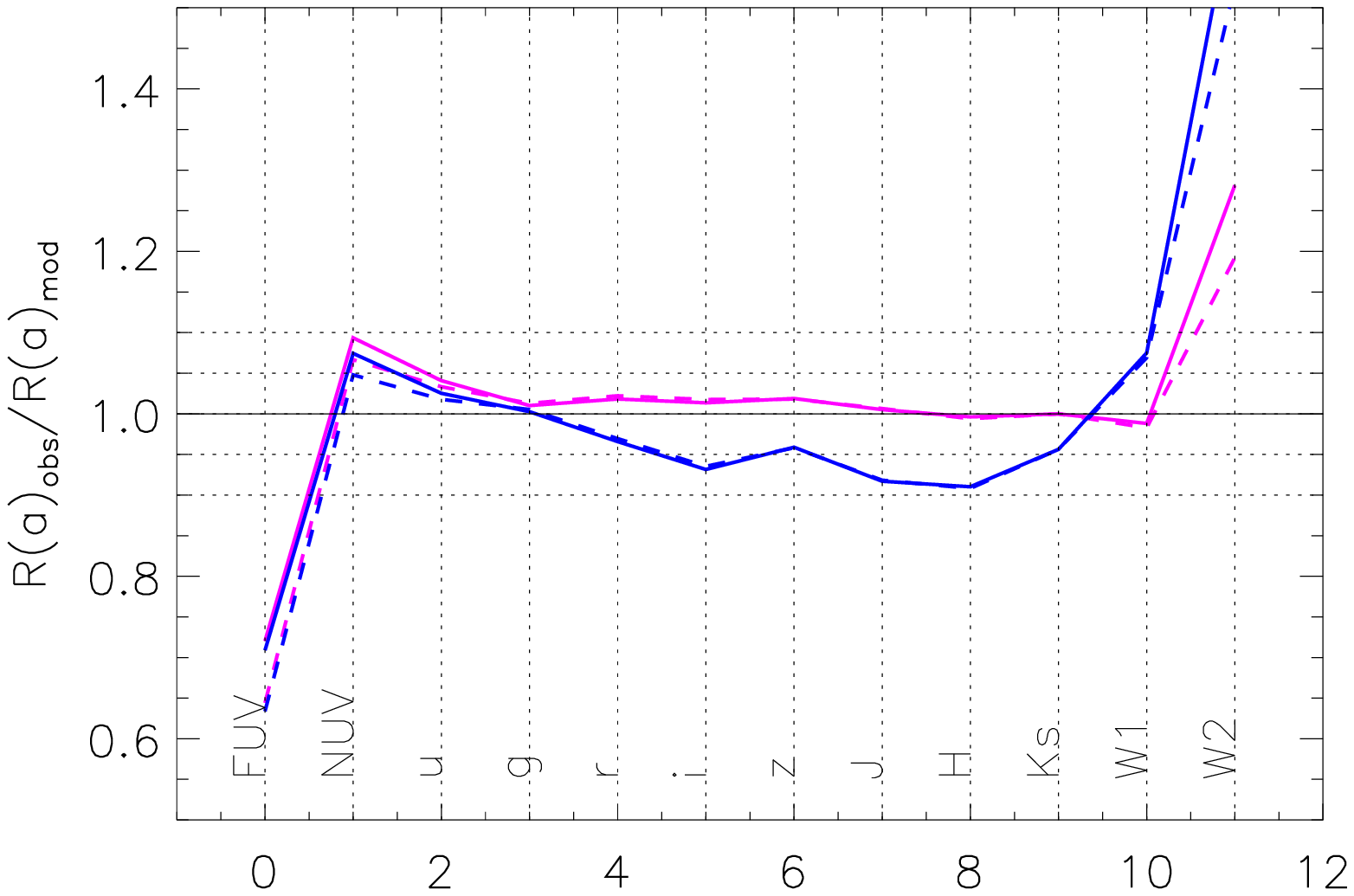}
\caption{
Comparison between the measured and predicted extinction coefficients using data from Table\,2.
The lines are similar to those in Fig.\,7.}
\label{}
\end{figure}

Fig.\,7 show comparison between the measured and predicted reddening coefficients, using data from Table\,1.
Fig.\,8 show comparison between the measured and predicted extinction coefficients, using data from Table\,2.
The purple solid and dashed lines represent the ratios of the measured coefficients in the 2nd and 3rd columns to
those predicted by the \rv~=~3.1 Fitzpatrick reddening law in the 4rd column, respectively.
The blue solid and dashed lines represent the ratios of the measured coefficients in the 2nd and 3rd columns to
those predicted by the \rv~=~3.1 CCM reddening law in the 5th column, respectively.
The cyan solid and dashed lines represent the ratios of the measured  coefficients in the 2nd and 3rd columns to
those predicted by the \rv~=~3.1 O'Donnell reddening law in the 6th column, respectively.
It is clearly seen that the two sets of coefficients derived from the data plotted in the left and right panels of
Figs.\,3--6 agree well with each other.
More importantly, the newly deduced reddening coefficients for the
$g-r, r-i, i-z, z-J, J-H, H-Ks$ and $Ks-W1$ colors agree well with 
the \rv~=~3.1 Fitzpatrick reddening law but disfavor the \rv~=~3.1 CCM and O'Donnell laws. 
This result suggests that the \rv~=~3.1 Fitzpatrick reddening law is preferred to 
the \rv~=~3.1 CCM and O'Donnell reddening laws.
However, the \rv~=~3.1 Fitzpatrick law seems to under-predict $R(NUV-u)$ and $R(u-g)$
values by about 12 per cent, indicating an update of the Fitzpatrick law in the UV is necessary.
Note that the observed $R(NUV-u)/R(g-r)$ and $R(u-g)/R(g-r)$ ratios agree better
with the CCM law than the Fitzpatrick law.

Both the \rv~=~3.1 Fitzpatrick and CCM laws fail to explain the observed $R(FUV-NUV)$ values.
The predicted $R(FUV-NUV)$ values are sensitive to temperature of the source spectrum used.
In this work, a 7,000\,K source spectrum is adopted to calculate the predicted reddening coefficients,
while the temperatures of most targets for the GALEX passbands are below 7,000\,K.
However, the predicted values increases as the temperature decreases, therefore 
a more representative source spectrum will increase the differences between 
the observed and predicted $R(FUV-NUV)$ values. The large differences could also be caused by
uncertainties of the source spectrum in the $FUV$ band. To further investigate the differences, 
a much hotter and larger sample is needed.
The Fitzpatrick and CCM reddening laws can not explain the observed $R(W1-W2)$ values either.
This is probably caused by the fact that neither the Fitzpatrick nor the CCM laws are
valid for the WISE bands, thus extrapolations have to be carried out to calculate
the predicted $R(W1-W2)$ values. Note that the observed small values of $R(W1-W2)$ are consistent with previous studies
of mid-IR extinction laws (e.g., Gao, Jiang \& Li 2009).

We have assumed \rv~=~3.1 in this work when calculating the predicted extinction
coefficients by the Fitzpatrick, CCM and O'Donnell reddening laws. Given that \rv~is
sensitive to the IR color excesses (Fitzpatrick 1999), we have performed a consistent check using
the data for the 2MASS passbands.  By minimizing the
differences between the observed reddening of $g-z$, $g-J$, $g-H$ and $g-Ks$
relative to $g-r$ and those predicted by Fitzpatrick reddening laws of different \rv,
we find the optimal value of \rv~for each star in the target sample for the 2MASS passbands.
A histogram of \rv~values thus obtained is plotted in Fig.\,9.
The median and mean values of \rv~ are 3.09 and 3.12, respectively, 
consistent with the value adopted in 
the current work as well as the average value for the Galactic diffuse interstellar medium.
If the mean value of \rv~is adopted, the differences between observations and the predictions of extinction laws will be slightly 
smaller by less than 1.0 per cent. However, the main results of this work are not affected.   

\begin{figure}
\includegraphics[width=90mm]{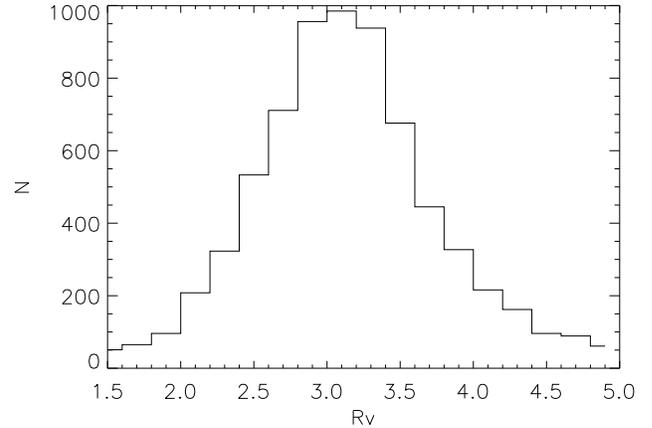}
\caption{Histogram of \rv~deduced using the sample for the 2MASS passbands.
The median and mean \rv~values of the sample are 3.09 and 3.12, respectively.
}
\label{}
\end{figure}

The SFD dust reddening map delivers the total amount of
reddening along a sightline, thus may have over-estimated the real value for a local
disk star.  However, the tight correlation between $E(g-r)$ and \ebv, as
shown in the bottom-right panel of Fig.\,3, suggests that most local disk stars
have been excluded from our samples and therefore this effect is unlikely to be import.
To quantify this potential effect, we select a subsample of stars of $|b| \ge 15\degr$
from the target sample for the SDSS passbands and re-fit the data. We find a new
value of 1.007 for $R(g-r)$, i.e. only 1.6 per cent larger than that from the original target sample.
The differences between the new and original values of $R(u-g)$ and $R(i-z)$ are even smaller,
confirming that this effect has not affected the main results of this work. 

In our analysis, we have not included the effects of possible variations of $R(g-r)$, 
which is sensitive to the normalization of reddening law, 
as a function of sky position and extinction.
Such variations do exist at the level of a few to ten per cent (S10; SF11). 
We have not considered the variations of extinction law as a function of 
sky position and extinction either. 
However, there is evidence indicating that the law is fairly universal over the SDSS footprint (SF11).
The reddening and extinction coefficients presented in the current work have been deduced largely based on targets within 
the SDSS DR7 footprint, and should be generally applied to this area.
They may not be suitable for other regions, e.g., the Galactic bulge and star-forming regions in the disk.

As noted earlier ({\S}\,2.1), values of \teff, \logg~and [Fe/H] adopted in the current
analysis are averaged values yielded by the NGS1, ki13, ANNRR and ANNSR methods of
SSPP, i.e. methods that do not rely on  photometric data.  
To investigate the possible systematic effects of inaccurate reddening corrections
on the SSPP adopted parameters, we have examined the differences between
the values of \teff~adopted in this work and those adopted by the SSPP as a function
of extinction.  We find that the systematic difference increases from below 30
K at \ebv~$\le 0.01$ to over 100 K at \ebv~$\sim 1.0$.  We suggest that
the reddening and extinction coefficients presented in this work should be adopted 
in the future versions of SSPP and other related studies. 

The "standard pair" technique to measure extinction to individual stars with photometric and spectroscopic observations,
to study the extinction laws and to detect and study the DIBs in the SDSS spectra 
can be easily extended to on-going and planned large scale spectroscopic surveys.
The Large Sky Area Multi-Object Fiber Spectroscopic Telescope (LAMOST; Wang et al. 1996; Su et al.
1998; Xing et al. 1998; Zhao 2000; Cui et al. 2004; Zhu et al. 2006;  c.f. http://www.lamost.org/website/en/)
is a 4-m class telescope that is capable of recording spectra of up to 4,000 objects 
simultaneously in a field of view of 5$^\circ$ in diameter.
Commencing in fall of 2012, the LAMOST Galactic surveys (Deng et al. 2012), 
including the LAMOST Digital Sky Survey of the Galactic Anti-center (DSS-GAC; Liu et al., in preparation), 
have started collecting  spectra for millions of stars down to $r \sim 18-19$.
The forth-coming next generation astrometric satellite Gaia (Perryman et al.
2001; Katz et al. 2004) will yield distances for one billion Galactic stars to $V \sim 20$.
By combining data from large spectroscopic surveys such as LAMOST and SDSS, with those from 
the astrometric survey of GAIA and from  
large photometric surveys from the UV to the IR including WISE, 2MASS, SDSS, GALEX, and 
the Xuyi Schmidt Telescope Photometric Survey of the Galactic Anticenter (XSTPS-GAC; Yuan et al. in preparation), 
Pan-STARRS (Kaiser et al. 2002) and LSST (Tyson 2002),
will enable us to carry out detailed studies of  
extinction and extinction laws, dust properties and distributions as 
well as DIBs and their carriers in a three-dimensional way.

\section{Summary}
With "star pairs" selected from SDSS spectroscopic archive,
combing the SDSS, GALEX, 2MASS and WISE photometry that ranges from the far UV to the mid-IR,
we have measured dust reddening in the $FUV-NUV, NUV-u, u-g, g-r, r-i, i-z, z-J, J-H, H-Ks, Ks-W1$ and $W1-W2$ colors for thousands of stars.
The measurements, together with \ebv~values from SFD, are used to deduce the observed, model-free reddening coefficients for those colors.
The new coefficients are compared to previous measurements in the literature and to the predictions of different dust reddening laws.
The results show that:
\begin{itemize}
\item The dust reddening map of SFD over-estimate \ebv~by about 14 per cent,
consistent with the earlier studies of S10 and SF11;
\item After taking into account the differences in reddening normalization, 
our newly deduced reddening coefficients for the
$FUV-NUV, NUV-u, u-g, g-r, r-i, i-z, z-J, J-H, H-Ks, Ks-W1$ and $W1-W2$ colors
differ by respectively 
$-$1640\%, 15.5\%, 12.6\%, $-$0.8\%, 3.4\%, $-$0.7\%, 3.5\%, 2.5\%, 1.4\%, 2.2\% and $-$50.7\% 
from the predictions of the \rv~=~3.1 Fitzpatrick reddening law, 
and by respectively $-$1730\%, 13.0\%, 8.1\%, 10.0\%, 8.0\%, $-$13.5\%,
$-$1.7\%, $-$6.7\%, $-$17.1\%, $-$17.7\% and $-$58.6\%
from the predictions of the \rv~=~3.1 CCM reddening law;
\item The new reddening coefficients for colors from $g-r$ to $W1-W2$ 
prefer the \rv~=~3.1 Fitzpatrick reddening law to the \rv~=~3.1 CCM and O'Donnell reddening laws. 
However, the Fitzpatrick law seems to under-predict $R(NUV-u)$ and $R(u-g)$
by about 12 per cent, indicating an update of the Fitzpatrick law in the UV is needed.
\end{itemize}

Using the extinction coefficient of $Ks$ band given by the \rv~=~3.1 Fitzpatrick law and the observed reddening 
coefficients presented in this work,
we have obtained new extinction coefficients for the $FUV, NUV, u, g, r, i, z, J, H, W1$ and $W2$ passbands.
We recommend that the new reddening and extinction coefficients should be generally used when performing reddening 
correction of Galactic stars with the SFD dust map in future. 
We stress however that the $FUV$- and $NUV$-band coefficients should be used with caution given their relatively large 
measurement uncertainties.

\vspace{7mm} \noindent {\bf Acknowledgments}{
We would like to thank the referee for his/her valuable comments, which helped improve the quality of the paper significantly.
This work has made use of data products from the Sloan Digital Sky Survey (SDSS),
Galaxy Evolution Explorer (GALEX), Two Micron All Sky Survey (2MASS), Wide-field Infrared Survey 
Explorer (WISE), SIMBAD database and NASA/IPAC Infrared Science Archive.
}

\label{lastpage}

\end{document}